\documentclass[preprint2]{aastex}       

\def\be{\begin{eqnarray}}
\def\ee{\end{eqnarray}}
\def\lsim{\;\raise0.3ex\hbox{$<$\kern-0.75em\raise-1.1ex\hbox{$\sim$}}\;}
\def\gsim{\;\raise0.3ex\hbox{$>$\kern-0.75em\raise-1.1ex\hbox{$\sim$}}\;}

\shorttitle{Neutral Heating of Shock Precursors }
\shortauthors{  }

\begin{document}

\title{Effects of Neutral Hydrogen on Cosmic Ray Precursors in Supernova Remnant Shock Waves}

\author{John C. Raymond,\altaffilmark{1}
        J. Vink,\altaffilmark{2}
        E.A. Helder,\altaffilmark{2}
        \&  A. de Laat\altaffilmark{2} }

\altaffiltext{1}{Harvard-Smithsonian Center for Astrophysics, 60 Garden St., Cambridge, MA
02138, USA; jraymond@cfa.harvard.edu}
\altaffiltext{2}{Astronomical Institute, Utrecht University, P.O. Box 80000, 3508TA Utrecht, The Netherlands}

\begin{abstract}
Many fast supernova remnant shocks show spectra dominated by Balmer lines.
The H$\alpha$ profiles have a narrow component explained by
direct excitations and a thermally Doppler broadened component
due to atoms that undergo charge exchange in the
post-shock region.
However, the standard model does not take into account
the cosmic-ray shock precursor, which compresses and accelerates
plasma ahead of the shock.
In strong precursors with sufficiently high densities, the processes of
charge exchange, excitation and ionization will affect the widths of 
both narrow and broad line components.  Moreover, the
difference in velocity between the neutrals and the precursor plasma
gives rise to frictional heating due to charge exchange and ionization in the
precursor.  In extreme cases, all neutrals can be ionized by the precursor.

In this paper we compute the ion and electron heating for a wide range of
shock parameters, along with the velocity distribution of the neutrals that
reach the shock.
Our calculations predict very large narrow component widths for some
shocks with efficient acceleration, along with changes in the broad- to-narrow
intensity ratio used as a diagnostic for the electron-ion temperature ratio.
Balmer lines may therefore provide a unique diagnostic of precursor properties.
We show that heating by neutrals in the precursor can account for the
observed H$\alpha$ narrow component widths, and that the acceleration efficiency is
modest in most Balmer line shocks observed thus far.

\end{abstract}


\section{Introduction}
\label{s_intro}

Cosmic rays are widely believed to originate in supernova remnant (SNR)
shock waves, because the cosmic-ray energy spectrum agrees with model predictions,
because power-law distributions of energetic electrons are seen in 
SNRs, and because the power required to maintain the cosmic ray population
could be supplied by about 10\% of kinetic energy of Galactic supernovae.
The standard theory for the process is Diffusive Shock Acceleration (DSA),
which is a first order Fermi process requiring that particles scatter 
between a gasdynamic subshock and plasma turbulence in a shock precursor.
Evidence for non-linear DSA comes
from curved synchrotron spectra
\citep{reynolds92,vink06,allen}, evidence for high compression factors
\citep{warren05,cassam08} and evidence for lower than expected downstream
temperatures \citep{hughes00,helder09}.
However, all this evidence is based on observation of downstream
properties. The effects of precursor physics on the H$\alpha$ emission
described here offer a direct probe of the properties
of the precursor.

A crucial parameter for these models is the diffusion coefficient
$\kappa$, as it determines the precursor scale length, which is typically
$\kappa$ divided by the
shock speed $V_S$.  Gas is compressed in the precursor and accelerated
to a fraction of the shock speed, and this compression is related
to $V_S$, to the efficiency of particle acceleration and to the
escape of energetic particles from the region \citep{bykov05, vink10}.
Neutrals can impede the acceleration
process by damping the turbulence needed to scatter particles
back to the shock.  However, \citet{drury} found that the
acceleration efficiency can be high as long as the density and
neutral fraction are not too large, though the maximum particle energy
is reduced.

One set of diagnostics for the physics of collisionless shocks is
based on the emission from particles in the narrow ionization zone
just behind a nonradiative shock \citep{ray91, heng10}.  In particular,
H$\alpha$ photons from a nonradiative shock in partly neutral gas originate
very close to the shock, and Coulomb collisions
do not have time to erase such signatures as unequal electron and ion
temperatures or non-Maxwellian velocity distributions \citep{laming96,
ghavam01, ray08, ray10}.  In the optical these shocks are seen as pure Balmer line
filaments whose profiles show a narrow component characteristic of
the pre-shock kinetic temperature and a broad component closely
related to the post-shock proton temperature \citep{cr78, heng10,
vanA}.  The intensity ratio of the broad and narrow components
is determined by the electron to ion temperature ratio at the shock
\citep{ghavam01, vanA, helder10t}.

The Balmer line profiles also contain signatures of shock precursors.
In general, the narrow component line widths are 40 to 50 $\rm km~s^{-1}$,
indicating temperatures around 40,000 K.  If that were the ambient ISM
temperature, there would be no neutrals to create the Balmer line
filament, so the width is interpreted as an indication of heating in
a narrow precursor too thin to completely ionize the hydrogen
\citep{smith94, hester94, lee07, sollerman}.  Faint emission
ahead of the sharp filament is interpreted as emission from the
compressed and heated precursor gas \citep{hester94, lee07, lee10}. 

This paper considers the role of neutrals in heating the precursor
plasma and computes the properties of precursor H$\alpha$ emission.
While cosmic ray pressure in the precursor can compress, heat and
accelerate ions and electrons by means of plasma turbulence and
magnetic fields, the neutrals only interact with the precursor
by means of collisions with protons and electrons.
If the density is very high, neutrals and protons are
tightly coupled by charge transfer.  In that case, the neutrals are compressed
along with the protons and adiabatically heated.  They also share
in any other heating of the protons, such as dissipation of Alfv\'{e}n
waves generated by cosmic-ray streaming.  On the other hand, if the
density is very low, neutrals pass through the precursor and the
shock without interacting at all, preserving their pre-shock
velocity distribution.

The intermediate case is more complex.  A shock that efficiently
accelerates cosmic rays is strongly modified, and gas reaches a
significant fraction of the shock speed in the precursor \citep{vladimirov, wagner09}.
If neutrals and ions are fairly well coupled, they can be described
as fluids whose relative speed gives a frictional heating similar
to that in C-shocks \citep{drainemckee}.  If a
neutral encounters this high speed compressed plasma without having
been brought gradually up to speed by many previous charge transfers,
it can be ionized and become a pickup ion \citep{ray08, ohira} like those observed in
the solar wind \citep{moebius85}.  It can then have an energy on the order
of 1 keV, which it can share with the other protons.  Electron heating
is more uncertain, but it can occur by means of Lower Hybrid waves
\citep{cairnszank}.  If the electrons are heated they
can excite and ionize H atoms, changing the H$\alpha$ profile and the
broad-to-narrow line ratio used as an electron temperature diagnostic \citep{ghavam01}.

In this paper we compute the proton, neutral and electron temperatures
in the precursors for a variety of parameters, along with the 
ionization and excitation of H atoms.  We consider the effects of
these processes on Balmer line diagnostics currently in use.
\citet{ohira} considered the effects of neutrals on the velocity structure of
the precursor, the compression ratio and the acceleration process.  They
found that the pickup ions can reduce the compression by the subshock and
enhance proton injection into the acceleration process.
\citet{morlino} self-consistently computed the particle acceleration
and heating due to neutrals, but within the fluid approximation for
both neutrals and ions.  In this paper we emphasize the effects on
the H$\alpha$ line profile.

\section{Model Calculations}
\label{s_model}

We parameterize
the precursor structure in a relatively simple manner.  We
assume that the precursor accelerates and compresses the interstellar
gas over a length scale $\kappa$/$V_S$, where $\kappa$ is the
diffusion coefficient for cosmic rays near the cutoff.
Effective cosmic-ray acceleration requires $\kappa$ on the
order of $10^{24}~\rm cm^2~s^{-1}$,  and estimates based on
the scales of H$\alpha$ precursors are 2 to 4$\times 10^{24}~\rm cm^2~s^{-1}$
\citep{lee07, lee10}.  We do not consider the second order
effects of momentum 
and energy deposition by the neutrals on
the precursor length scale.

We assume an exponential form, so that the compression is given by

\begin{equation}
  \chi ~=~ 1 + (\chi_1 -1) ~~ e^{(x  V_S / \kappa)}
\end{equation}

\noindent
where x is negative ahead of the shock and $\chi_1$ is 
the compression ratio just upstream of the subshock.
It is related to the fractional pressure of cosmic rays behind the 
shock, w=$P_{CR} / (P_G + P_{CR})$, by equation 9 of \citet{vink10}.  We simplify this
equation with the
assumption that for w$<$0.8 the compression in the gas subshock equals 4, so that

\begin{equation}
\chi_1~=~ (1-w/4)/(1-w)
\end{equation}

\noindent
Mass conservation implies velocities $V = V_S / \chi$ in the frame of the shock.

To compute the proton and electron temperatures we include
adiabatic compression, Coulomb energy transfer between
protons and electrons, energy losses due to ionization and
excitation of Hydrogen, and heating terms.

We assume that any neutral that interacts with the plasma at
position $x_i$ joins the proton flow at that position.  If the interaction 
was charge transfer, a new neutral is formed with the bulk speed and
thermal speed of the protons at $x_i$. Thus the neutrals arriving at $x_i$
are those that last went through charge transfer at all upstream positions
$x_j$, and they have the speeds, $v_j$, of the plasma at $x_j$.  Each ionization
of a neutral from $x_j$ at $x_i$ deposits energy
$0.5 m_p (v_j-v_i)^2$.  We assume that the energy is quickly
thermalized among the protons, unlike \citet{ohira}, who assumed
a pickup ion velocity distribution.  The thermalization time scale is
very uncertain, because full kinetic calculations have not been carried out.
However, in the highly
turbulent precursor there are many wave modes besides Alfv\'{e}n
waves that can thermalize the protons, in particular those
associated with bump-on-tail, mirror and firehose-like instabilities
\citep{winske, gary, sagdeev}.  We also ignore heating 
due to Alfv\'{e}n wave damping or shocks excited by the cosmic-ray 
pressure gradient in order to isolate the effect of the neutrals. 
Therefore, we compute a lower limit to the heating.

For electrons, we follow \citet{cairnszank},
who found that ionization of fast neutrals forms a ring beam, in which all 
the particles gyrate around the magnetic field with the same speed but 
different phases.  The ring beam is unstable, and provided that the beam 
velocity (in this case the relative velocity of bins i and j) is less
than 5 times the Alfv\'{e}n speed, it transfers a significant
amount of heat to electrons via Lower Hybrid waves.  We follow
\citet{cairnszank} in taking this fraction to be 10\%.  Again,
to isolate the effects of neutrals we ignore any heating of
electrons by Lower Hybrid waves generated by cosmic-ray
streaming \citep{ghavam07, rakow08}.

Charge transfer rates are taken from \citep{schultz} using the
quadrature sum of the thermal speed and the ion-neutral relative speed.  Ionization
and excitation rates are computed from cross sections from \citet{janev} by 
integrating over the electron velocity distribution
including the relative electron-neutral flow speed.

 \section{Results}
 \label{s_results}

Figure 1 shows a set of models for a shock speed of
2000 $\rm km~s^{-1}$ with $\kappa = 2.0 \times 10^{24}~\rm cm^2~s^{-1}$,
a pre-shock density of 0.2 $\rm cm^{-3}$ and a neutral fraction of
0.2.  The four models have ratios of cosmic-ray partial pressure to total
pressure behind the subshock, w, of 0.1, 0.3, 0.5 and 0.7.  The compression
ratios just ahead of the subshock are 1.0833, 1.3214, 1.75 and 2.75.
In the high $V_S$, high efficiency models the neutrals are
not compressed to this level, because of collisional ionization
and because some pass through without charge transfer.
The protons and electrons are strongly heated in the more efficient
models, but the electrons are much cooler than the protons.  The
drop in heating just before the subshock in the 70\% efficient model
results from the reduced number of neutrals.

Figure 2 shows the velocity distributions of the neutrals perpendicular
to the shock just before the subshock. 
Note that the w=50\% model shows a narrow component
due to neutrals that last experience charge transfer far upstream, along with
a broader component of particles that undergo charge transfer close to
the subshock.

Figure 3 shows a grid of models of the neutral velocity distribution at 
the shock for a range of shock speeds and cosmic-ray
partial pressures.  The panels show the FWHM measured directly from the 
computed velocity distribution and the kurtosis, which would be 3.0 for
a Gaussian distribution.  Kurtosis is a problematic statistical moment
for real data because it is sensitive to noise far from the
line center and the choice of background level.  However, for the
theoretical profiles computed here it highlights cases in which
some neutrals undergo charge transfer close to the subshock and others
do not.  We also show the fraction of incident neutrals that survive up to
the subshock and the average number of excitations to the n=3 level per
incident H atom.  Not all of these excitations will result in H$\alpha$ photons because
some Ly$\beta$ photons escape, but this is a convenient comparison to
the 0.2 to 0.25 H$\alpha$ photons per H atom produced in the post-shock region.

 \section{Discussion} 
 \label{s_discussion}

The interaction of neutral hydrogen with the ionized plasma in cosmic-ray
precursors described above offers an important tool to measure
the properties of cosmic-ray precursors.  
The outcome of DSA is very much influenced by
physical processes in the precursor, which are not well determined.
For example, non-adiabatic heating and magnetic field amplification due the presence
of cosmic rays tend to decrease the overall compression factor
from $\chi_{12}>>20$ \citep[e.g.][]{berezhko,blasi05} to
$7 \lesssim  \chi_{12}\lesssim 15$ \citep{vladimirov, caprioli}. In addition, if the Alfv\'en
waves in the precursor have some drift velocity this will affect the
cosmic-ray pressure profile \citep{zirakashvili08}, which
limits the escape of energy from the shock region. A lower 
energy escape automatically implies a
lower downstream cosmic-ray pressure \citep{vink10}.

As shown here, neutrals will influence the physics of the precursor.
\citet{morlino} treated the neutrals as a fluid coupled to the ions
by charge transfer for a unified treatment
of the heating and dynamics of the precursor, but the fluid approximation
is only appropriate if neutrals and ions are coupled fairly well.
They obtained a FWHM of 46 $\rm km~s^{-1}$ for
the H$\alpha$ line in a 2000 $\rm km~s^{-1}$ shock with modest efficiency,
a pre-shock density of 1 $\rm cm^{-3}$ and 50\% neutral fraction.  For
similar parameters we 
find a non-Maxwellian profile with smaller FWHM and broader wings.
 
Neutrals can also damp plasma waves, which limits the efficiency of cosmic-ray
acceleration.  This damping is caused by the central processes
described above: charge exchange and ionization. 
\citet{drury} found that the maximum particle energy, and therefore the 
maximum acceleration efficiency, is considerably
higher than suggested by \citet{drainemckee}.  In addition, the heating
due to neutrals penetrating the precursor is a form of non-adiabatic heating.
Energy dissipated in the precursor
limits the amount of free energy available for shock acceleration.  If the 
neutrals ionized in the precursor behave as pickup ions rather than 
thermalizing with the protons, the injection efficiency and particle
spectrum will be affected \citep{ohira}.  In any case, the heating of 
electrons in the precursor is poorly know, and that will stongly affect the
ionization and excitation of H atoms, which in turn will affect the
intensity ratio of the broad and narrow components as well as the 
narrow component line width. 

The physics of neutral-ion coupling means these processes
are not only sensitive to cosmic-ray pressure and the structure
of the precursor, but also to the pre-shock density and neutral fraction.
For example, the protons and neutrals in the Cygnus Loop nonradiative shocks
\citep{salvesen} are tightly coupled, and they behave nearly 
adiabatically, while the pre-shock density in SN1006 is so low \citep{acero07} 
that neutrals pass straight through it. This may explain the narrow line 
width seen by \citet{sollerman} in SN\,1006
in comparison to broader narrow lines observed for other young SNRs.
The number of charge transfer events for an average neutral in the precursor
can be estimated roughly as

\begin{equation}
N_{\rm CT} = nL\sigma \chi_1 / V_S ,
\end{equation}

\noindent
where L is the precursor length scale, and the charge transfer cross section, $\sigma$,
declines slowly with velocity below about 2000 km s$^{-1}$, then very rapidly.

Comparing our results to observations, it is obvious that the observed narrow-line
H$\alpha$ widths are in general smaller than predicted by our calculations for efficient
shock acceleration (i.e., $w>0.5$). This may indicate that none of the shocks investigated so far
accelerate particles efficiently. However, more work is needed before such a conclusion
can be drawn, as the line width depends also on pre-shock density and shock velocity.
For very high shock velocities combined with low densities the neutrals hardly interact in the precursor,
leading to narrow line widths. Such may be the case for the northeastern region of RCW 86, for which
\citet{helder09} reported a high cosmic-ray acceleration efficiency 
\citep[$w \gtrsim 0.5$, see also][]{vink10}. For this
region the pre-shock density may be as low as $n\lesssim 0.1$ \citep{vink06}, which, combined
with the high velocity ($V_s\gtrsim 3000$~km/s), gives few interactions in the precursor and
widths $\lesssim 100$~km/s.  Note that this is smaller than could be measured given the moderate
spectral resolution of the measurement.  

Perhaps the most striking result from these calculations is that for efficient shocks near 1000
$\rm km~s^{-1}$ a substantial number of H$\alpha$ photons will be produced in the precursor.  These
will usually be included in the narrow component, potentially affecting the electron temperature
estimate based on the broad-to-narrow intensity ratio \citep{ghavam01}.  Narrow component
emission from the precursor could explain the broad-to-narrow intensity ratios that
cannot be fit by models of post-shock emission \citep{vanA, rakow09}.  In extreme cases, 
emission from the precursor might also contribute to the broad component, possibly accounting for the
non-Maxwellian profile seen in Tycho's SNR \citep{ray10} and generally leading to an underestimate
of the shock speed.  Both of these conclusions depend on the diffusion coefficient and the electron
heating, however.

Other H$\alpha$ line measurements show that the narrow lines are broader than one might expect
for temperatures of typical HII regions, but smaller than 50~km/s \citep{sollerman}.
(Not all of the narrow line emission comes from the precursor, but the narrow line emission 
downstream is determined by the velocity distribution in the precursor.)
Another effect of charge exchange in the precursor is that neutrals enter the downstream shock
region with a velocity offset with respect to the local interstellar medium, as seen in
Tycho's supernova remnant (Lee et al., 2007). For shocks observed
face on this should produce a narrow line offset, which for the combined front and back side
of the remnant should lead to two narrow lines. The spectra of several LMC remnants \citep{smith94}
do not show such an effect. For one of the remnants in this set, SNR 0509-67.5, the
cosmic-ray acceleration efficiency was estimated to be $w\approx 0.2$ \citep{helder10}.

We note that several improvements should be made to the calculations
presented here.  Additional heating due to wave dissipation can heat
the protons, resulting in larger narrow component line widths, or it can
heat electrons, increasing the H$\alpha$ narrow component intensity and 
reducing the number of neutrals that reach the shock, especially if the
electron velocity distribution is non-Maxwellian \citep{laming07}.  In addition, ionization and excitation by proton and helium ion
impact are important at high relative velocities \citep{laming96}, and at high shock speeds
the velocity distributions of particles are anisotropic \citep{hengmccray, heng07, vanA}.
Amplification of the magnetic field may also be important, and radiative transfer calculations in the Ly$\beta$ line
must be done to compute the H$\alpha$ emission.  We plan to address these issues in future work.

\acknowledgements

This work was carried out while JCR was visiting the Astronomical Institute Utrecht 
as Minnaert Professor.  It was supported by NASA grant GO-11184.01-A-R to
the Smithsonian Astrophysical Observatory.  JV and EH are supported by the
VIDI grant awarded to JV by the Netherlands Science Foundation (NWO).

\bibliographystyle{apj}

\onecolumn

\begin{figure}
\includegraphics[width=0.95\textwidth]{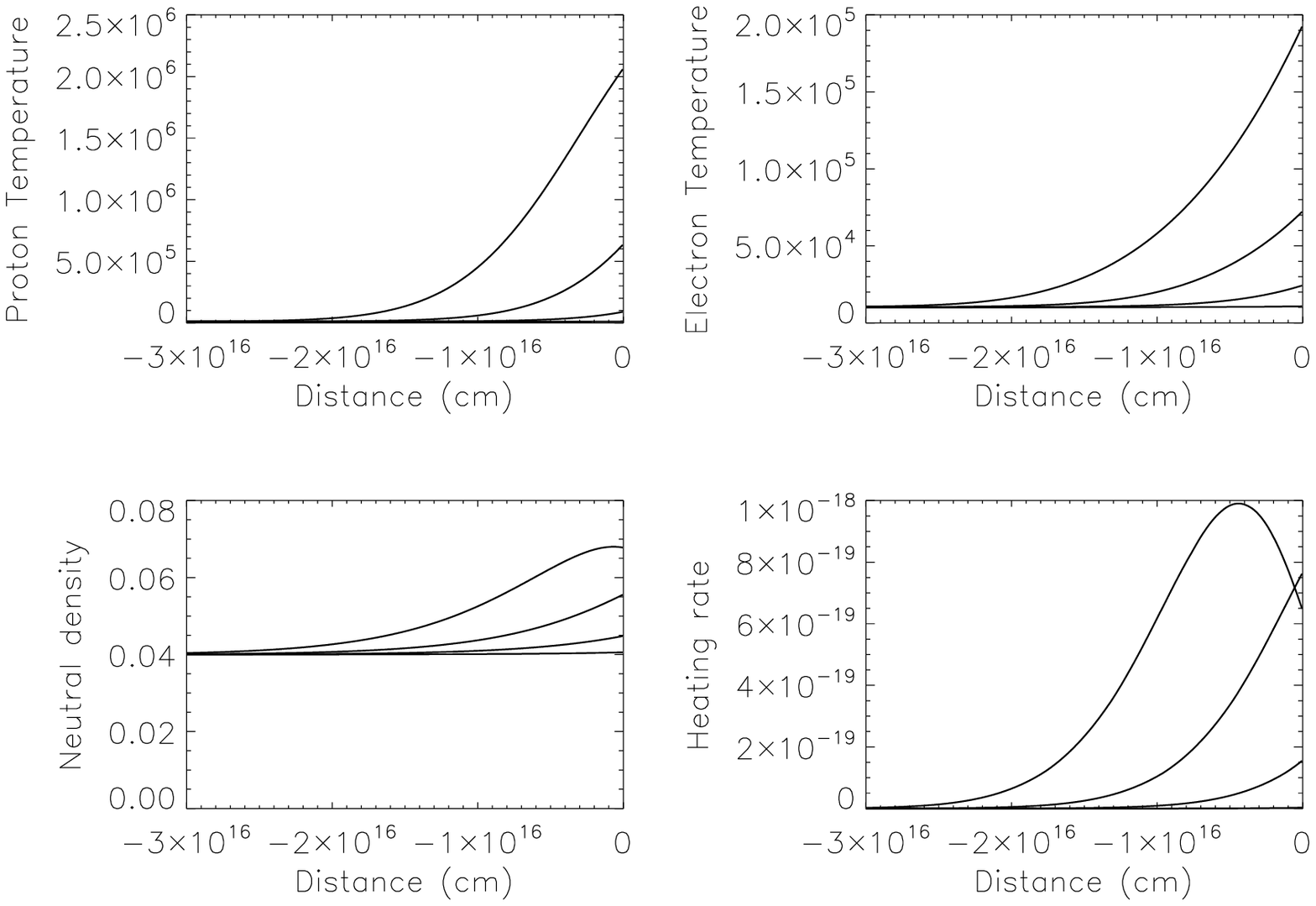}
\caption{Plots of a) proton temperature, b) electron
temperature, c) neutral density and d) heating rate
for for models having $\kappa=2.0 \times 10^{24}~\rm cm^2~s^{-1}$,
pre-shock density = 0.2 $\rm cm^{-3}$ and neutral fraction 0.2.  The models
assume a post-shock cosmic-ray pressure of 10\%, 30\%, 50\% and 70\%
of the total pressure, with the 10\% curves at the bottom and the 70\%
curves at the top in all four plots.
}
\label{f_model4}
\end{figure}

\begin{figure}
\includegraphics[width=0.45\textwidth]{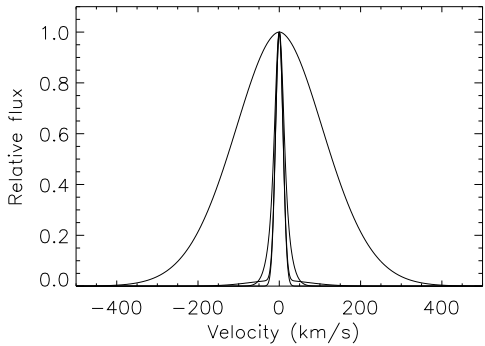}
\caption{Velocity distributions perpendicular to the flow
direction at the gasdynamic subshock for the four models shown in
Figure 1.  Note the narrow central component and broader wings in
the model with 50\% cosmic-ray pressure and the large width
predicted by the 70\% cosmic-ray pressure model.
}
\label{f_model4pro}
\end{figure}

\begin{figure}
\includegraphics[width=0.95\textwidth]{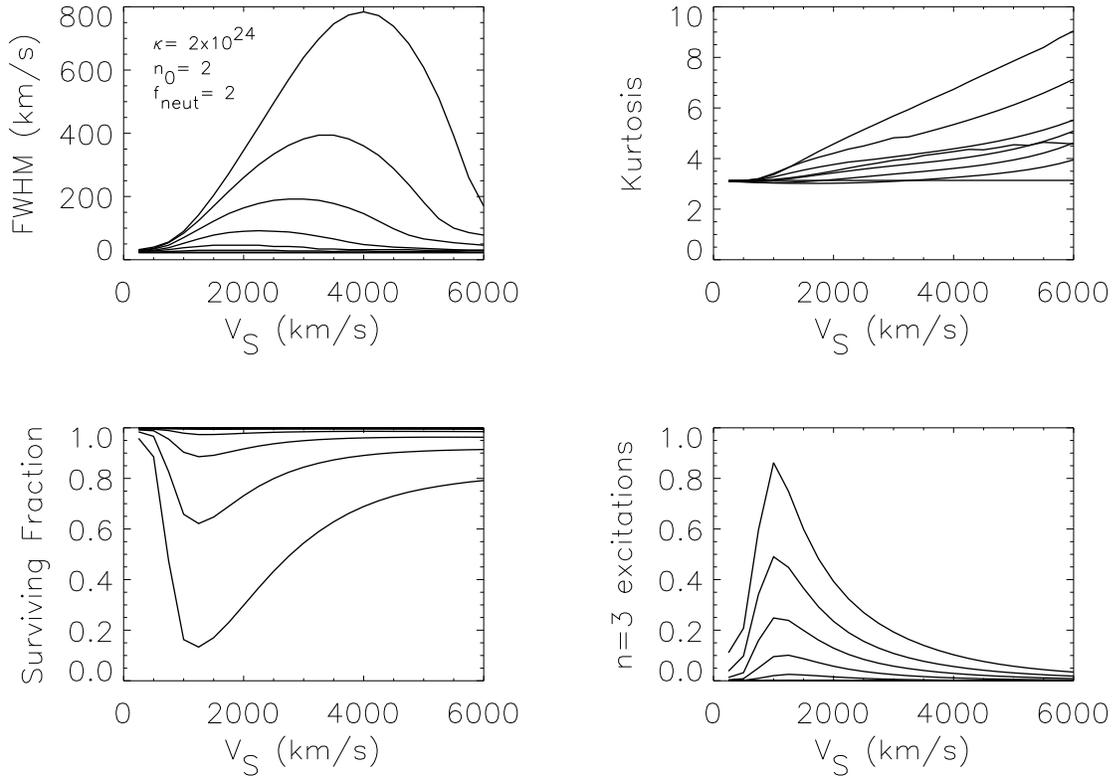}
\caption{Model grid for $\kappa = 2.0\times 10^{24}$, a pre-shock
density of 0.2 $\rm cm^{-3}$ and neutral fraction of 0.2.  Models
are shown for $P_{CR} / (P_{CR} + P_{G})$ = 0.1 to 0.8 at the shock
front.  The panels show the FWHM and kurtosis of neutrals that
reach the subshock, the
fraction of neutrals that reach the subshock and the number of
excitations to n=3 in the precursor per incident neutral hydrogen atom.  As in Figure
1, the 80\% models are the extreme cases.}
\label{f_grid}
\end{figure}

\end{document}